\documentclass[PRB,twocolumn,showpacs,preprintnumbers,superscriptaddress,prb]{revtex4}
\usepackage{amsmath}
\usepackage{dcolumn}
\usepackage{bm}
\usepackage{longtable}
\usepackage{graphicx}

\begin{document}

\title{Anisotropy in the magnetic and electrical transport properties of Fe$%
_{1-x}$Cr$_{x}$Sb$_{2}$}
\author{Rongwei Hu}
\affiliation{Condensed Matter Physics and Materials Science Department, Brookhaven
National Laboratory, Upton, NY 11973}
\affiliation{Department of Physics, Brown University, Providence, RI 02912}
\author{V. F. Mitrovi{\'c}}
\affiliation{Department of Physics, Brown University, Providence, RI 02912}
\author{C. Petrovi{c}$^{1}$}
\date{\today}

\begin{abstract}
We have investigated anisotropy in magnetic and electrical transport
properties of Fe$_{1-x}$Cr$_{x}$Sb$_{2}$ ($0\leq x\leq 1$) single crystals.
The magnetic ground state of the system evolves from paramagnetic to
antiferromagnetic with gradual substitution of Fe with Cr. Anisotropy in
electrical transport diminishes with increased Cr substitution and fades
away by $x=0.5$. We find that the variable range hopping (VRH) conduction
mechanism dominates at low temperatures for $0.4\leq x\leq 0.75$.
\end{abstract}

\pacs{71.20.Nr, 71.28.+d, 75.30.-m, 75.50.Pp}
\maketitle

\section{Introduction}

The magnetic resemblance of FeSb$_{2}$ to FeSi suggests that FeSb$_{2}$
could be another model system to study the possible applicability of the
Kondo insulator scenario to transition metal compounds with \textit{3d}
electrons.\cite{Petrovic1}$^{,}$\cite{Petrovic2}$^{,}$\cite{Perucchi}$^{,}$%
\cite{TRice} The magnetic susceptibility of FeSb$_{2}$ is nearly temperature
independent at low temperatures, followed by diamagnetic to paramagnetic
crossover around 100 K and thermally activated behavior at high temperatures.%
\cite{Rongwei1} Another striking similarity to FeSi comes from the result of 
\textsl{ab initio} LDA+\textit{U} electronic calculations that show near
degeneracy of small gap semiconductor and metallic ferromagnetic state.\cite%
{TRice} Electrical transport in FeSb$_{2}$ is rather anisotropic and nearly
quasi one-dimensional. Whereas resistivity along the $\widehat{a}$- and $%
\widehat{c}$- axes is semiconducting showing activated temperature
dependence below room temperature, the $\widehat{b}$- axis resistivity is
metallic above 40 K and semiconducting below that temperature.\cite%
{Petrovic1} Optical conductivity measurements showed a true insulating
state, zero Drude weight of $\sigma (\omega )$ at low frequencies and an
anisotropic energy gap E$_{g}$ in the spectral range between 100 - 350 cm$%
^{-1}$ in overall agreement with resistivity measurements. Thermal
excitations of charge carriers through E$_{g}$ in conventional MIT
transitions produce redistribution of the spectral weight just above the
gap. However, a full recovery of spectral weight in FeSb$_{2}$ occurs above
1 eV, suggesting contributions of larger energy scales.\cite{Perucchi}
Electron doping on Fe site induced a weak ferromagnetic metallic state and
colossal magnetoresistance in Fe$_{1-x}$Co$_{x}$Sb$_{2}$. Thus other doping
studies in this system are of particular interest.\cite{Rongwei1}$^{,}$\cite%
{Rongwei2}

Hole doping, on the other hand, induces a heavy fermion metallic state in
FeSb$_{2-x}$Sn$_{x}$\cite{Danes}, analogous to the case in FeSi$_{1-x}$Al$%
_{x}$.\cite{DiTusa} To better understand the influence of the hole doping on
the ground state of FeSb$_{2}$, we investigate the effects of Cr
substitution on Fe site of this material. The significance of Cr
substitution is that holes are introduced in narrow energy bands with\textit{%
\ 3d} character, unlike in the case of Co and Sn substitution. CrSb$_{2}$ is
an antiferromagnetic semiconductor that crystallizes into the marcasite
structure with the same space group \textit{Pnnm}.\cite{Goodenough} Neutron
diffraction studies of CrSb$_{2}$ indicate a high spin \textit{d}$^{\mathit{2%
}}$ configuration, an antiferromagnetic ground state with Neel temperature T$%
_{N}=273\pm 2\,\mathrm{K}$, and a magnetic moment per Cr of 1.94 $\mu _{B}$.%
\cite{Holseth} In this work we examine the magnetic and electrical
properties of Fe$_{1-x}$Cr$_{x}$Sb$_{2}$ (0$\leq x\leq 1$) and discuss the
possible conduction mechanism of this binary system.

\section{Experimental Method}

\begin{figure}[b]
\centerline{\includegraphics[scale=0.62]{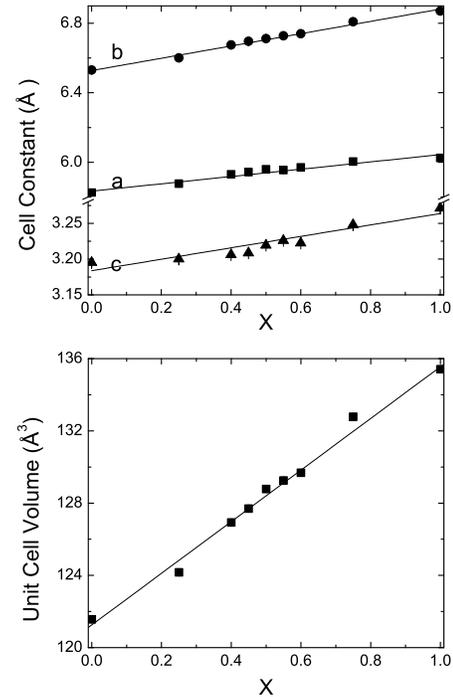}} 
\vspace*{-0.3cm}
\caption{{\protect\small Lattice constants and unit cell volume of Fe$_{1-x}$%
Cr$_{x}$Sb$_{2}$ \textit{versus} nominal Cr concentration $x$.}}
\end{figure}

The samples investigated in our experiments were single crystals grown from
high temperature melt.\cite{Canfield}$^{,}$\cite{Fisk} Powder X-ray
diffraction (XRD) spectra of the ground samples were taken with Cu K$%
_{\alpha }$ radiation ($\lambda =1.5418$ \AA ) using a Rigaku Miniflex X-ray
machine. The lattice parameters were obtained by fitting the XRD spectra
using the Rietica software.\cite{Hunter} Single crystals were oriented using
a Laue Camera and were polished into rectangular bars along specific
crystalline axes. Thin Pt wires were attached to electrical contacts made
with Epotek H20E silver epoxy for a standard 4-wire resistance measurement.
Sample dimensions were measured with an optical microscope Nikon SMZ-800
with 10 $\mu $m resolution, and the average values were used to estimate a
geometric factor. Magnetization, resistivity and heat capacity measurements
were carried out in a Quantum Design MPMS-5 and a PPMS-9 for temperatures
from 1.8 K$\ $to 350 K.

\section{Results and discussion\protect\bigskip}

The lattice parameters of doped samples from the powder X-ray diffraction
spectra are shown in Fig. 1. Linear dependence on Cr concentration in
accordance with Vegard's law demonstrates that Cr uniformly substitutes Fe
in the entire doping range. All axes increase linearly with increasing $x$
and the lattice constants of the end members are consistent with the
reported values, FeSb$_{2}$ ($a=5.8253(2)$ \AA , $b=6.5313(2)$ \AA , $%
c=3.1952(2)$ \AA ) and CrSb$_{2}$ ($a=6.0250(1)$ \AA , $b=6.8708(1)$ \AA , $%
c=3.2711(1)$ \AA ).\cite{Petrovic1}$^{,}$\cite{Kjekshus1}$^{,}$\cite{Harada}

\begin{figure}[t]
\centerline{\includegraphics[scale=0.87]{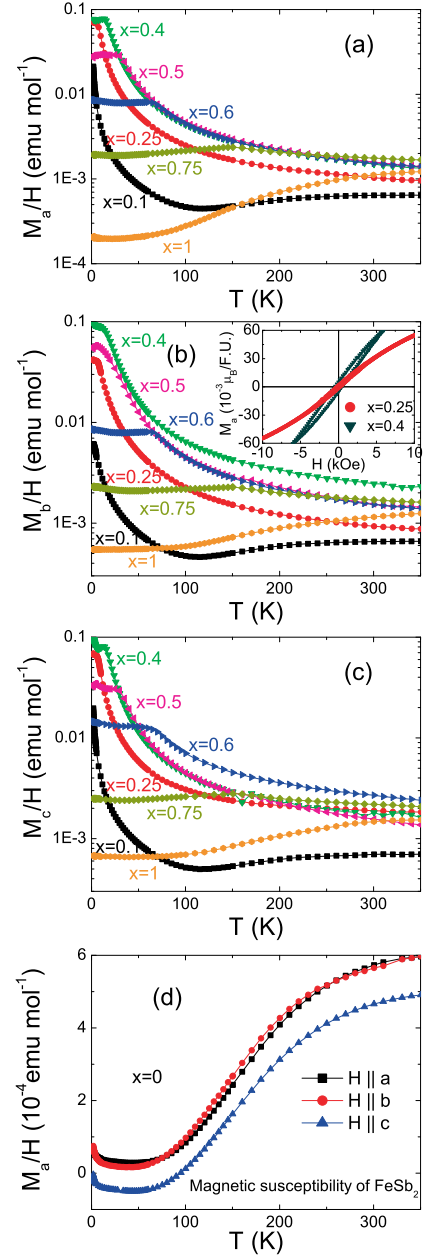}} 
\vspace*{-0.2cm}
\caption{{\protect\small \ \textbf{(a)}-\textbf{(c)}Temperature dependance
of }${\protect\small M/H}$ {\protect\small for field of $H=1000$ Oe applied
along 3 crystalline axes. Inset shows the hysteresis loops for }$%
{\protect\small 0.25}\leq x\leq 0.4$. {\protect\small \textbf{(d)} $M/H$ 
\textit{versus} temperature ($T$) of pure FeSb}$_{2}.$}
\end{figure}

\begin{table*}[tbh]
\caption{Parameters of the fits to the Curie-Weiss law of the high
temperature $M/H$ and semiconducting energy gaps of the resistivity.}
\label{Tb1}%
\begin{tabular}{p{2.2cm}p{2.2cm}p{2.2cm}p{2.2cm}p{2.2cm}p{2.2cm}p{2.2cm}}
\hline\hline
$x$ & $\mu _{eff}(\mu _{B})$ & $\Theta (K)$ & $\chi _{0}(emu/mol)$ & $%
T_{C}/T_{N}(K)$ & $Ms(10^{-3}\mu _{B})$ & $\Delta _{\rho }(K)$ \\ \hline
$0.10$ & $1.08(6)$ & $22(8)$ & $1.3\times 10^{-4}$ &  &  & $302(9)$ \\ 
$0.25$ & $1.13(5)$ & $26(4)$ & $8.2\times 10^{-4}$ & $6.6(5)$ & $45(1)$ & $%
321(2)$ \\ 
$0.40$ & $1.85(6)$ & $18(3)$ & $3.8\times 10^{-4}$ & $14(1)$ & $72(1)$ & $%
434(8)$ \\ 
$0.50$ & $1.88(2)$ & $2.5(8)$ & $2.2\times 10^{-4}$ & $25$ &  & $425(1)$ \\ 
$0.60$ & $1.82(4)$ & $-12(3)$ & $6.3\times 10^{-4}$ & $65(2)$ &  & $397(6)$
\\ 
$0.75$ & $1.44(3)$ & $-18(1)$ & $12.2\times 10^{-4}$ & $150(5)$ &  & $251(4)$
\\ \hline\hline
\end{tabular}%
\end{table*}

The magnetic susceptibility along the three crystal axes of Fe$_{1-x}$Cr$%
_{x} $Sb$_{2}$ ($0\leq x\leq 1$) in temperature range 1.8 to 350 K is shown
in Fig. 2. FeSb$_{2}$ exhibits a diamagnetic to paramagnetic crossover at
about 100 K for field applied along the $\widehat{c}$ axis. The temperature
dependence of the susceptibility along the $\widehat{a}$- and $\widehat{b}$-
axes is virtually identical: nearly temperature independent at low
temperatures and thermally activated at higher temperatures. The magnetic
susceptibility of CrSb$_{2}$ is consistent with magnetic ordering below T$%
_{N}$=275 $K$. A free ion model Eq.(1) proposed by Jaccarino applied to FeSb$%
_{2}$ gives a spin gap $\Delta _{\chi }=527\pm 4$ $K$.\cite{Jaccarino} 
\begin{eqnarray*}
\chi _{FI} &=&Ng^{2}\mu _{B}^{2}\frac{J(J+1)}{3k_{B}T}\frac{2J+1}{2J+1+\exp
(\Delta _{\chi }/k_{B}T)}+\chi _{0}\text{ }(1) \\
\chi _{CW} &=&\frac{N\mu _{eff}^{2}}{3k_{B}(T-\Theta )}\text{ \ \ \ }\qquad
\qquad \qquad (2)
\end{eqnarray*}%
For $x=0.1$, the temperature dependence of susceptibility is comprised of a
reduced thermally activated term, Eq.(1), with gap value $\Delta _{\chi
}=441\pm 8$ $K$, and a low-temperature Curie-Weiss term, Eq.(2), with $\mu
_{eff}=$ $1.08\pm $ $0.06$ $\mu _{B}$ due to Cr substitution. With further
Cr substitution, long range antiferromagnetic order is induced for $x\geq
0.25$ and the Neel temperature gradually increases toward T$_{N}$ =275 K, of
CrSb$_{2}$ $(x=1)$. The high temperature effective magnetic moment is
somewhat lower than expected, $\mu _{eff}$(Cr$^{4+}$) = $2$ $\mu _{B}$ for
the end concentrations of $x=0.1$ and $x=0.75$, as summarized in Table I.
This deviation could arise due to contributions of the thermally activated
FeSb$_{2}$-like Pauli susceptibility for low Cr doping and to temperature
limitations of our magnetic susceptibility measurement, where $\chi (T)$
curves were fitted in close proximity to the antiferromagnetic transition
for $x=0.75$.

\begin{figure}[b]
\centerline{\includegraphics[scale=0.75]{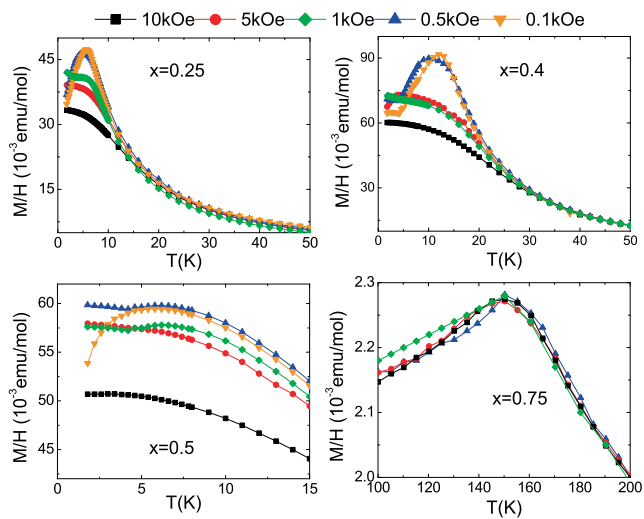}} 
\vspace*{-0.2cm}
\caption{{\protect\small Field dependance of the magnetic transitions for }$%
{\protect\small 0.25}\leq x\leq 0.75$}
\end{figure}

Parameters deduced from the fit to the Eq. (1) model are listed in Table I.
Clearly, the Curie-Weiss temperature changes sign from positive to negative
indicating that the magnetic coupling evolves from ferromagnetic to
antiferromagnetic. However, the field dependence of the transition, as
depicted in Fig. 3, reveals a canted antiferromagnetic structure for $%
0.25\leq x\leq 0.45$. In this doping range, in a predominantly
antiferromagnetic phase a small spontaneous magnetization is present due to
a slight deviation from a strictly antiparallel arrangement or to parasitic
ferromagnetism. Since there is an inversion symmetry at the Fe site in the
Pnnm space group of FeSb$_{2}$, we can exclude the presence of
Dzayloshinsky-Moriya type of interaction.\cite{Lebech} The seemingly
ferromagnetic tail is sensitive to the applied field, changing to a
characteristic antiferromagnetic peak in lower fields. Magnetism in Fe$%
_{1-x} $Cr$_{x}$Sb$_{2}$ is most likely attributable to the ordering of Cr
atoms since the electronic configuration of Fe in FeSb$_{2}$ is the
nonmagnetic \textit{3d}$^{4}$. Assuming that the magnetization $M(H)$
consists of a saturation moment($M_{s}$) and paramagnetic component ($\chi H$%
), we obtain the canting angle $\phi $ of $2.2\sim 4.1%
{{}^\circ}%
$. The angle is defined as $2\sin ^{-1}(M_{S}/2M_{\max })$ where $M_{\max
}=2\,\mu _{B}$ is the magnetic moment of the Cr$^{4+}$. The saturation
moment $M_{S}\sim 0.09\mu _{B}/\mathrm{Cr}$ accounts for the ordering of
4.5\% of Cr atoms. Hysteresis loops with coercive fields of $H_{C}=(230\sim
1100)$ $Oe$ and remnant fields of $H_{R}=(10\sim 63)$ $Oe$ $emu/mol$ were
observed with field applied along all three crystalline axes in this doping
range. With further Cr substitution for $0.6\leq x\leq 1$, the system has
the antiferromagnetic structure of CrSb$_{2}$, consistent with the results
of Kjekshus \textit{et al}..\cite{Kjekshus1}

\begin{figure}[tb]
\centerline{\includegraphics[scale=1.6]{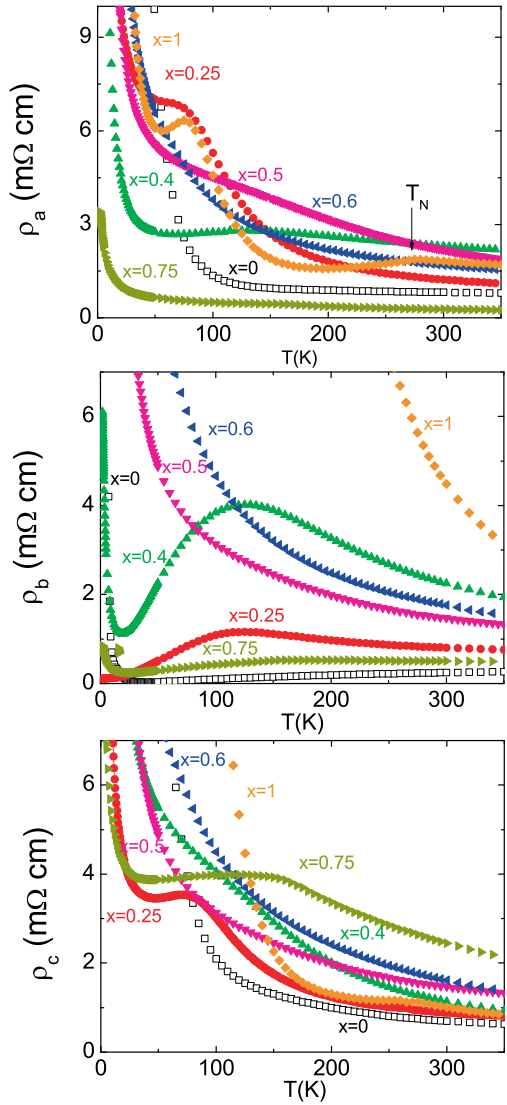}} 
\vspace*{-0.2cm}
\caption{{\protect\small Resistivity of Fe$_{1-x}$Cr$_{x}$Sb$_{2}$ for
current applied along three crystalline axes. Arrow indicates the Neel
temperatures.}}
\end{figure}

The temperature dependence of the resistivity of Fe$_{1-x}$Cr$_{x}$Sb$_{2}$
for current applied along individual axes of the crystal structure is shown
in Fig. 4. Electrical transport along the $\widehat{b}$ -\ axis shows a
substantial metallic region for $0\leq x\leq 0.25$. The region of metallic
resistivity gradually shifts down in temperature with increasing $x$,
yielding a metallic state at the lowest measured temperature for $x=0.25$.
At high temperatures, however, the region of metallic resistivity is
progressively reduced with the increasing $x$, giving way to semiconducting
behavior above 150 K. Above 150 K activated temperature dependence of
resistivity is isotropic for all $x\geq 0.1$, except for $\rho _{b}$ of $%
x=0.75$.

The evolution of the intrinsic band semiconducting energy gap is listed in
Table I. The gap values increase in nonmonotonic fashion from $\Delta _{\rho
}=225\pm 5$ $K$ for $x=0$ to $\Delta _{\rho }=834\pm 3$ $K$ for $x=1$. For $%
x=0.4$, the gap value reaches its local maximum. Semiconducting energy gap
for $x=0.75$ deviates from monotonic behavior possibly due to semimetallic
transport observed for $\widehat{b}$-axis resistivity at high temperatures.
There is a loss of spin disorder scattering at the T$_{N}$ for $0.25\leq
x\leq 0.1$, in good agreement with magnetization measurements.

\begin{figure}[t]
\centerline{\includegraphics[scale=0.8]{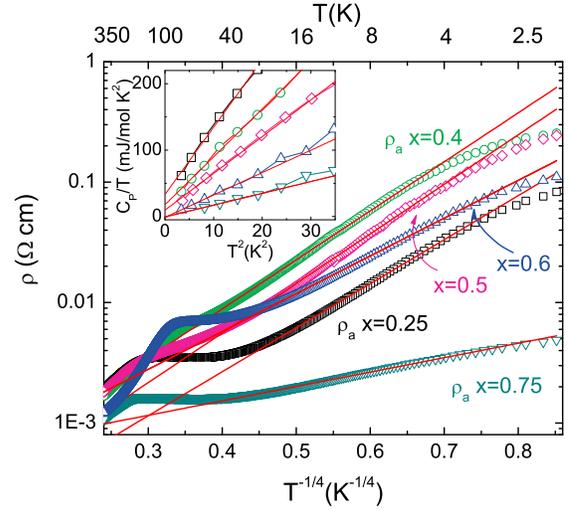}} 
\vspace*{-0.2cm}
\caption{ $\protect\rho $ \textit{versus} $T^{-1/4}$ curves for $0.25\leq
x\leq 0.75$ on a semilog scale. Inset: C$_{p}/T$ \textit{versus} $T^{2}$ for 
$T < 6$ K. }
\end{figure}

The resistivity data for an intermediate range of $x$, $0.25\leq x\leq 0.75$%
, is depicted in Fig. 5. Low temperature resistivity is described by $\rho
=\rho _{0}\exp (T_{0}/T)^{1/4}$, implying a variable range hopping (VRH)
conduction mechanism. The characteristic temperatures $T_{0}$ and the
corresponding localization lengths $\xi $ are summarized in Table II. The
length is given by $\xi ^{3}=19/[k_{B}T_{0}N(E_{F})]$\cite{Friedman}, where
the density of states at the Fermi level $N(E_{F})$ of $3\times 10^{37}\,%
\mathrm{J}^{-1}\mathrm{cm}^{-3}$ is estimated using low temperature specific
heat measurements (Fig. 5 inset).

\begin{table}[b]
\caption{T$_{0}$ and $\protect\xi $ of variable range hopping and Sommerfeld
coefficient from specific heat measurements.}%
\begin{tabular}{p{2cm}p{1.2cm}p{1.2cm}p{1.2cm}p{1.2cm}p{1.2cm}}
\hline\hline
$x$ & $0.25$ & $0.4$ & $0.5$ & $0.6$ & $0.75$ \\ \hline
$T_{0}(K)$ & $2723$ & $4657$ & $7063$ & $7750$ & $57$ \\ 
$\xi (\mathring{A})$ & $215$ & $193$ & $198$ & $254$ & $1758$ \\ 
$\gamma (mJ/molK^{2})$ & $33.1$ & $17.7$ & $10.7$ & $2.1$ & $0.1$ \\ 
\hline\hline
\end{tabular}%
\end{table}

\begin{figure}[h]
\centerline{\includegraphics[scale=0.8]{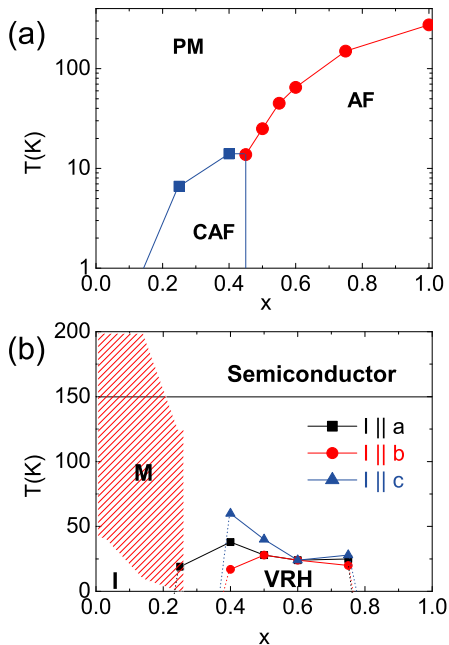}} 
\vspace*{-0.4cm}
\caption{{\protect\small Phase diagram of Fe}$_{1-x}${\protect\small Cr}$%
_{x} ${\protect\small Sb}$_{2}$, compiled from measurements of magnetic
properties \textbf{(a)} and from electrical transport \textbf{(b)}. }
\end{figure}

The rich phase diagrams of the electronic system in the Fe$_{1-x}$Cr$_{x}$Sb$%
_{2}$ alloys are displayed in Fig. 6. With the increase of $x$, the
paramagnetic ground state evolves toward an antiferromagnetic one, see Fig.
6(a). In the antiferromagnetic part of the phase diagram we distinguish a
canted antiferromagnetic (CAF) region for $0.2\leq x\leq 0.45$. The phase
diagram compiled from the electrical transport properties is shown in Fig.
6(b). The $\widehat{b}$ - axis transport is metallic for $0\leq x\leq 0.25$.
In addition, there are two semiconducting regions. The system is band gap
semiconductor for $0.25<x$ at high temperature. The variable range hopping
(VRH) mechanism is present for intermediate alloys for $0.25\leq x\leq 0.75$%
. The low temperature insulating region in FeSb$_{2}$ diminishes with
increasing $x$ in Fe$_{1-x}$Cr$_{x}$Sb$_{2}$ and disappears by \thinspace $%
x=0.25$.

FeSb$_{2}$ and CrSb$_{2}$ crystallize in an orthorhombic marcasite \textit{%
Pnnm} structure, whereas CoSb$_{2}$ has a pseudomarcasite monoclinic \textit{%
P21/c} symmetry induced by an increase of $\widehat{a}$ - axis, doubling of
the $\widehat{c}$ - axis and a small monoclinic angle $\beta =90.4^{\circ }$%
. FeSb$_{2}$, CoSb$_{2}$ and CrSb$_{2}$ are all semiconducting materials and
continuously miscible on their respective binary alloy phase diagrams. Thus,
it is possible to control a rich formation of ground states by carrier
concentration. Both Fe$_{1-x}$Co$_{x}$Sb$_{2}$ and Fe$_{1-x}$Cr$_{x}$Sb$_{2}$
electronic systems are noteworthy in that although the two end members are
semiconducting, metallic regions exist in their binary alloy systems. In
contrast to the intermediate Fe$_{1-x}$Co$_{x}$Sb$_{2}$ system, where a
metallic region appears with as small an electron doping as $x=0.05$, in Fe$%
_{1-x}$Cr$_{x}$Sb$_{2}$ ground state metallicity is observed only for $%
x=0.25 $. This is consistent with LDA+U band structure calculations.\cite%
{TRice} The density of states (DOS) above the Fermi level is largely
determined by the occupancy of $3d$ orbitals. Electron doping readily moves
the Fermi level into the large peak just above the Fermi energy, in contrast
to hole doping which moves the Fermi level to the depleted region of the DOS
toward the valence band.

\section{Conclusion}

We have studied anisotropy in the magnetic and electrical transport
properties of Fe$_{1-x}$Cr$_{x}$Sb$_{2}$. The substitution of Cr for Fe
induced a rich phase diagram. Magnetically ordered ground states are formed
for $0.25\leq x\leq 1$. The canted antiferromagnetism of intermediate alloys
for $x\leq 0.5$ evolves into the antiferromagnetic state of CrSb$_{2}$ with
a Neel temperature gradually increasing up to T$_{N}$=275 K for $x=1$. The
material is semiconducting, with a metallic region for $\widehat{b}$ - axis
transport in the paramagnetic region for $0\leq x\leq 0.25$. The
isostructural Fe$_{1-x}$Cr$_{x}$Sb$_{2}$ alloys are an interesting
playground for studying intermediate electronic states between small gap,
nearly magnetic (or \textquotedblleft Kondo") semiconductor ($x=0$) and
antiferromagnetic semiconductor ($x=1$). Further optical and X-ray
scattering measurements could be useful to study this problem in more detail.

\section{Acknowledgments}

We thank W.-G. Yin and Myron Strongin for useful communication and critical
reading. This work was carried out at the Brookhaven National Laboratory,
which is operated for the U.S. Department of Energy by Brookhaven Science
Associates (DE-Ac02-98CH10886). This work was supported by the Office of
Basic Energy Sciences of the U.S. Department of Energy. This work was also
supported in part by the National Science Foundation DMR-0547938 (V. F. M.).

\end{document}